\documentclass[journal]{IEEEtran}

\usepackage[utf8]{inputenc}
\usepackage{amsmath}
\usepackage{cite}
\usepackage{multicol}
\usepackage{graphicx}
\usepackage{array}
\usepackage{multirow}
\usepackage{graphicx}
\usepackage{enumitem}
\usepackage{color}
\usepackage{nomencl}
\usepackage{etoolbox}
\usepackage{amssymb}
\usepackage{comment}
\usepackage{mathtools}
\usepackage{bbold}
\usepackage{yfonts}
\usepackage{soul}
\usepackage{hyperref}
\usepackage{fnpct}
\ifCLASSOPTIONcompsoc \usepackage[caption=false,font=normalsize,labelfon
t=sf,textfont=sf]{subfig}
\else
\usepackage[caption=false,font=footnotesize]{subfi g}
\fi
\usepackage{tikz}
\usetikzlibrary{decorations.pathreplacing}
\usetikzlibrary{arrows,shapes,positioning}
\usetikzlibrary{patterns}
\usepackage{pgfplots}

\title{Optimizing Phase Allocation in Unbalanced \\ Power Distribution Networks using a\\ Linearized DistFlow Formulation
\thanks{This work was supported by the Swiss National Science Foundation (SNSF), Postdoc Mobility under Grant \# 217874. }
}

\author{
\IEEEauthorblockN{Rahul K. Gupta$^*$, Daniel K. Molzahn$^\S$}\\
\IEEEauthorblockA{$^*$School of Electrical Engineering \& Computer Science\\
Washington State University, Pullman, 99163, WA, USA }\\
\IEEEauthorblockA{$^\S$School of Electrical and Computer Engineering \\
Georgia Institute of Technology, Atlanta, 30313, GA, WSA}\\
\IEEEauthorblockA{\{rahul.k.gupta@wsu, molzahn@gatech\}.edu}
}

\makeatletter

\begin{document}
\maketitle

\begin{abstract}
Power distribution networks, especially in North America, are often unbalanced but are designed to keep unbalance levels within the limits specified by IEEE, IEC, and NEMA standards. However, rapid integration of unbalanced devices, such as electric vehicle (EV) chargers and single-phase solar plants, can exacerbate these imbalances. This increase can trigger protection devices, increase losses, and potentially damage devices. To address this issue, phase swapping (or phase allocation) has been proposed. Existing approaches predominantly rely on heuristic methods. In this work, we develop a mixed integer linear programming (MILP) approach for phase allocation. Our approach uses linearized DistFlow equations to represent the distribution network and incorporates a phase consistency constraint, enforced with binary variables, to ensure that downstream phase configurations align with upstream configurations. We validate the proposed approach on multiple benchmark test cases and demonstrate that it effectively improves network balance, as quantified by various metrics.
\end{abstract}


\section{Introduction}
The nature of power injections into power distribution networks is progressively changing due to the rapid growth of single-phase solar plants and residential electric vehicle chargers. This shift may lead to a higher level of unbalance across phases, which could result in triggered protection devices, increased network losses, and damage to certain devices, such as distribution transformers. Unbalanced phases lead to inefficient network utilization, as one phase in a three-phase network may be more stressed than the other two.
To address these issues, organizations such as IEEE \cite{cooper1987ieee}, IEC \cite{compatibility2009environment}, and NEMA \cite{nema} have defined different unbalance metrics for distribution network operations~\cite{girigoudar2019naps}. Unbalance limits based on these metrics are essential to ensure the safe and reliable operation of household appliances and power equipments.

In the existing literature, several methods are proposed to address the issue of voltage imbalance, which can be broadly categorized into three types. The first method involves investing in new devices, such as static or dynamic VAR compensators \cite{xu2010voltage}, which can be costly. The second method uses active and reactive power control from existing or newly installed solar inverters and energy storage systems \cite{weckx2014reducing, weckx2015load, geth2015balanced, sun2015phase,girigoudar2023}. Many studies suggest regulating the active and reactive powers from solar PV inverters~\cite{girigoudar2023,gupta2024improving}. This strategy may require investments in the communication and situational awareness infrastructure for real-time control~\cite{gupta2020grid}.

The third approach is the phase switching or swapping strategy \cite{zhu1998phase, khodr2006optimal, zhang2024phase}, which involves reassigning certain loads to different phases at a bus to balance loads and reduce voltage unbalance. There are also hybrid approaches that combine phase switching with the use of static VAR compensators \cite{liu2020unbalance}. Utilities often employ these methods manually, where maintenance crews travel to sites and physically switch loads. This process is costly, requiring extensive planning and potentially costing tens of thousands of dollars per phase swap \cite{wang2013phase}. These balancing maneuvers are typically performed periodically, as load characteristics vary with seasonal changes. 

Given the high cost of manual phase balancing, it is essential to optimize the phase configuration selected for implementation. Recently, several studies have proposed phase-swapping algorithms, although this remains an under-researched area. For example, \cite{wang2013phase} developed and compared various heuristic schemes for phase swapping, and \cite{gandomkar2004phase} suggested using a genetic algorithm. Other works proposed using an immune algorithm \cite{huang2008three} and simulated annealing \cite{zhu1999phase} for phase balancing. However, these methods are computationally intensive, making it challenging to obtain solutions for large networks, and they may converge only to locally optimal solutions. The work in \cite{zhu1998phase} proposed a mixed integer linear programming (MILP), but does not model the grid constraints explicitly. 

This paper proposes a new approach to solving the phase allocation/swapping/balancing problem using a MILP formulation. We employ a linearized grid model, specifically an unbalanced adaptation of the Linearized DistFlow model (Lin3DistFlow) \cite{arnold2016optimal}, which enables a linear formulation of the problem. Binary variables are associated with each phase, determining which phase is active or inactive based on the nodal injections at each bus. Additionally, we introduce a specific constraint to ensure phase consistency, which means that downstream lines or nodes cannot have a phase if it is absent in the upstream system. This is achieved by pre-computing the child nodes of each bus and applying a hard constraint through the binary variables. The proposed MILP formulation offers distinct advantages over heuristic methods: it guarantees global optimality, obtains reproducible solutions, and benefits from robust support by commercial solvers such as Gurobi.

The paper is organized as follows. Section~\ref{sec:distribution_grid_model} reviews the linearized grid model used for phase voltage modeling and introduces the phase consistency model. Section~\ref{sec:phase_allocation}  presents the proposed phase allocation problem using the linearized grid model. Section~\ref{sec:numerical_sim} provides the test cases and validation results, and finally, Section~\ref{sec:conclusion} concludes the work.
\section{Distribution Grid Modeling}
\label{sec:distribution_grid_model}
This section introduces the linearized grid model and the phase consistency constraint that will be used in the next section for the proposed optimal phase allocation problem.
\subsection{Linearized DistFlow Model (Lin3DistFlow)}
We model grid constraints using the Linearized DistFlow (\mbox{Lin3DistFlow}) model, derived from the branch-flow model known as the \mbox{``DistFlow''} equations, originally proposed in \cite{baran1989network}. 
\begin{figure}[t]
\vspace{-1em}
    \centering
    \includegraphics[width=0.8\linewidth]{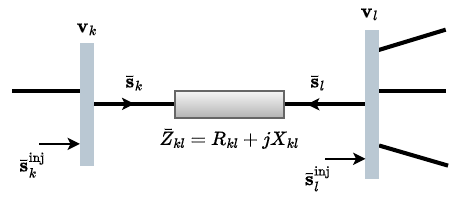}
    \vspace{-1em}
    \caption{Branch flow model in radial grids.}
    \label{fig:BFM}
\end{figure}
A schematic representation of the branch-flow model in radial grids is shown in Fig.~\ref{fig:BFM}.
Let the symbol $\mathbf{v}_k = [v_{k,a}~v_{k,b}~v_{k,c}]^\top$ be the \emph{squared} voltage magnitudes at bus~$k$. 
Let $\bar{\mathbf{s}}_{lk} = {\mathbf{p}}_{lk} + j{\mathbf{q}}_{lk}$ denote the complex\footnote{Symbols with $\bar{.}$ denote a complex quantity.} power flow on the line from bus $l$ to $k$, where $\mathbf{p}_{lk} = [p_{lk,a}~p_{lk,b}~p_{lk,c}]^\top$ and $\mathbf{q}_{lk} = [q_{lk,a}~q_{lk,b}~q_{lk,c}]^\top$ are the corresponding active and reactive power flows. Let $\bar{\mathbf{i}}_{lk}$ denote the current flow on the line from bus $l$ to $k$ and $\bar{Z}_{kl}$ be the line impedance matrix. Let $\bar{\mathbf{s}}^\text{inj}_l =\mathbf{p}^\text{inj}_l + j \mathbf{q}^\text{inj}_l $ be the complex power injection at bus $l$, where $\mathbf{p}^\text{inj}_l = [{p}^\text{inj}_{l,a}~{p}^\text{inj}_{l,b}~{p}^\text{inj}_{l,c}]^\top$ and $\mathbf{q}^\text{inj}_l = [{q}^\text{inj}_{l,a}~{q}^\text{inj}_{l,b}~{q}^\text{inj}_{l,c}]^\top$.

With the above notation, the DistFlow equations are:
\begin{subequations}
\begin{align}
    &  \mathbf{v}_k = \mathbf{v}_l - 2\Re(\bar{\mathbf{s}}_{lk}\bar{Z}_{kl}^{*}) +  \bar{Z}_{kl}\bar{\mathbf{i}}_{kl}\bar{\mathbf{i}}_{kl}^{*}\bar{Z}_{kl}^{*}, \\
    & \sum_{k:(l,k) \in \mathcal{L}} (\bar{\mathbf{s}}_{kl} + \bar{Z}_{kl}\bar{\mathbf{i}}_{kl}\bar{\mathbf{i}}_{kl}^*) - \bar{\mathbf{s}}_{lk} + \bar{\mathbf{s}}^\text{inj}_l = 0,
\end{align}
\end{subequations}
where $(\cdot)^{*}$ is the complex conjugate of a complex quantity.
LinDist3Flow, the linearized approximation of the DistFlow model, neglects the loss term ($\bar{\mathbf{i}}_{kl}\bar{\mathbf{i}}_{kl}^{*}$) to obtain:
\begin{subequations}
\begin{align}
    & \mathbf{v}_k \approx \mathbf{v}_l - 2\Re(\bar{\mathbf{s}}_{lk}\bar{Z}_{kl}^{*}), \label{eq:lindist_vmodel}\\
    & \sum_{k:(l,k) \in \mathcal{L}} \bar{\mathbf{s}}_{kl} - \bar{\mathbf{s}}_{lk} + \bar{\mathbf{s}}^\text{inj}_l  \approx 0. \label{eq:power_balance}
\end{align}

As illustrated in \cite{arnold2016optimal}, the expression \eqref{eq:lindist_vmodel} is simplified as: 
\begin{align}
    \mathbf{v}_k \approx \mathbf{v}_l + \mathbb{M}_{lk}^P \mathbf{p}_{lk} + \mathbb{M}_{lk}^Q \mathbf{q}_{lk},
\end{align}
where $\mathbb{M}_{lk}^P, \mathbb{M}_{lk}^Q$ are defined using per-phase resistances $r^{\phi,\phi'}_{lk}$ and reactances $x^{\phi,\phi'}_{lk}$ and ${\phi,\phi'} \in \{a,b,c\}$ denote the phase indices. The matrices $\mathbb{M}_{lk}^P, \mathbb{M}_{lk}^Q$ are
\begin{align}
    & \mathbb{M}_{lk}^P = \begin{bmatrix}
        -2r^{aa}_{lk} & r^{ab}_{lk} - \sqrt{3} x^{ab}_{lk} & r^{ac}_{lk} + \sqrt{3} x^{ac}_{lk}\\
    r^{ba}_{lk} + \sqrt{3} x^{ba}_{lk} & - 2r^{bb}_{lk} & r^{bc}_{lk} - \sqrt{3} x^{bc}_{lk}\\
        r^{ca}_{lk} - \sqrt{3} x^{ca}_{lk} & r^{cb}_{lk} - \sqrt{3} x^{cb}_{lk} & - 2r^{cc}_{lk}\\
    \end{bmatrix}, 
\end{align}
\begin{align}
     &   \mathbb{M}_{lk}^Q = \begin{bmatrix}
        -2x^{aa}_{lk} & x^{ab}_{lk} + \sqrt{3} r^{ab}_{lk} & x^{ac}_{lk} - \sqrt{3} r^{ac}_{lk}\\
    x^{ba}_{lk} - \sqrt{3} r^{ba}_{lk} & - 2x^{bb}_{lk} & x^{bc}_{lk} + \sqrt{3} r^{bc}_{lk}\\
        x^{ca}_{lk} + \sqrt{3} r^{ca}_{lk} & x^{cb}_{lk} + \sqrt{3} r^{cb}_{lk} & - 2x^{cc}_{lk}\\
    \end{bmatrix}.
\end{align}
\end{subequations}
\subsection{Phase Consistency}
\label{sec:phase_consistency}
\begin{figure}[!htbp]
    \vspace{-1em}
    \centering
    \includegraphics[width=\linewidth]{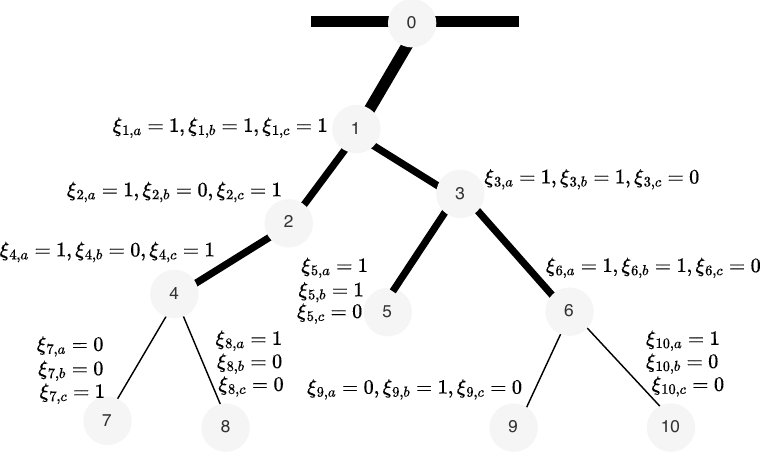}
    \caption{Example of a phase-consistent assignment problem.}
    \label{fig:phase_example}
\end{figure}

North American distribution grids typically consist of a mix of single-phase, two-phase, and three-phase systems. Consequently, when designing phase configurations, modelers need to ensure phase consistency across the network. In this context, phase consistency means that if a downstream line connects to an upstream line, the downstream line's phases cannot include a phase that is not included in the upstream line.
This concept is illustrated schematically in Fig.~\ref{fig:phase_example}. In this example, the line between nodes 0 and 1 carries all three phases, while the lines between nodes 1-2, 1-3, 2-4, 3-5 and 3-6 each carry two phases, and other connections use single-phase lines. The status of the phases is indicated by the symbols $\xi_{n,a}, \xi_{n,b}, \xi_{n,c}$ for node $n$.
In the existing literature, such constraints are often addressed with heuristic or iterative methods. Phase consistency is typically checked after the initial optimization (e.g., \cite{saha2019framework, postigo2020phase}), and, if inconsistencies are found, additional constraints are added and the optimization is rerun, repeating until a phase-consistent network is obtained. 

\vspace{-0.5em}
In this paper, we introduce a formulation that models phase consistency without the need for an iterative process. The symbols $\xi_{n,a}, \xi_{n,b}, \xi_{n,c}$ denote binary variables associated with each phase for a particular node $n$; these variables take the phase values available in the upstream line. The following constraints enforce phase consistency for this particular network:
\begin{subequations}
\vspace{-1em}
\begin{align}
   &  \xi_4 \geq \xi_7,\quad \xi_4 \geq \xi_8, \quad \xi_2 \geq \xi_4, \quad \xi_1 \geq \xi_2, \quad \xi_1 \geq \xi_3, \\
   &  \xi_0 \geq \xi_1, \quad \xi_6 \geq \xi_9, \quad \xi_6 \geq \xi_{10}, \quad \xi_3 \geq \xi_5, \quad \xi_3 \geq\xi_6.
\end{align}
\end{subequations}
In a general way, the constraint can be imposed as follows:
\begin{align}
    \xi_m \geq \xi_n, \quad n \in m_\text{children},
\end{align}
where $m_\text{children}$ denotes the immediate children nodes (downstream nodes) of node $m$.
\section{Phase Allocation Problem}
\label{sec:phase_allocation}
This section introduces the proposed phase allocation problem, which aims to optimize phase configurations throughout the network to achieve balanced voltages.

The objective of the phase allocation problem is to determine an optimal phase configuration by minimizing the difference between the voltage magnitudes among different phases at a node (i.e., the phase unbalance) and the number of phases per node, with the former weighted by a parameter $\alpha \leq 1$. The objective function minimizes the difference between each phase's voltage magnitudes from the mean of the voltage magnitudes at that node. Defining the mean of the squared voltage magnitude for a node $n$ as $\underline{v}_n = \frac{1}{3}(\sum_{\phi, \in \{a,b,c\}}v_{n,\phi})$, 
the phase balancing problem is formulated as
\begin{subequations}
\label{eq:phase_allocation}
\begin{align}
\begin{aligned}
    & \text{minimize}~\sum_{n\in\mathcal{N}}\sum_{\phi, \in \{a,b,c\}} (\alpha|\underline{v}_{n} - v_{n,\phi}| +   \xi_{n,\phi})
    \end{aligned}
    \label{eq:obj}
\end{align}
subject to:
\begin{align}
\label{eq:v_model}
        \begin{bmatrix}
            v_{k,a}\\
            v_{k,b}\\
            v_{k,c}
        \end{bmatrix} = 
        \begin{bmatrix}
            v_{l,a}\\
            v_{l,b}\\
            v_{l,c}
        \end{bmatrix} + \mathbb{M}_{lk}^P \begin{bmatrix}
            p_{lk,a}\\
            p_{lk,b}\\
            p_{lk,c}
        \end{bmatrix}
        + \mathbb{M}_{lk}^Q\begin{bmatrix}
            q_{lk,a}\\
            q_{lk,b}\\
            q_{lk,c}
        \end{bmatrix},
\end{align}
\begin{align}
     & 0\leq p^\text{inj}_{n,\phi} \leq \bar{p}_{n,\phi}~\xi_{n,\phi}, \quad \forall n\in\mathcal{N}, \phi \in \{a,b,c\}, \label{eq:p_inj_cons}\\
     & 0\leq q^\text{inj}_{n,\phi} \leq \bar{q}_{n,\phi}\quad\xi_{n,\phi}, \quad \forall n\in\mathcal{N}, \phi \in \{a,b,c\}, \label{eq:q_inj_cons}\\
     & 0\leq q^\text{inj}_{n,\phi} \leq {p}^\text{inj}_{n,\phi}, \quad \forall n\in\mathcal{N}, \phi \in \{a,b,c\}, \label{eq:pf_constraint}\\
     &  \sum_\phi p_{n,\phi} = \sum_\phi \hat{p}_{n,\phi} , \quad \forall n \in \mathcal{N}, \label{eq:psum_inj_cons}\\
      &  \sum_\phi q_{n,\phi} = \sum_\phi \hat{q}_{n,\phi} , \quad \forall n \in \mathcal{N}, \label{eq:qsum_inj_cons}\\
     & \sum_\phi \xi_{n,\phi} \leq |\mathcal{P}_n|, \quad \forall n \in \mathcal{N}, \label{eq:phase_constraint}\\
     & \xi_{m,\phi} \geq \xi_{n,\phi}, \quad n \in m_\text{child}, \quad \xi_{m,\phi} \in  \{0,1\}, \label{eq:phase_consistency}\\
    & \sum_{k:(l,k) \in \mathcal{L}} p_{lk, \phi} + {p}^\text{inj}_{l,\phi}  = 0, \quad \forall l\in\mathcal{N}, \phi \in \{a,b,c\} \label{eq:power_balance_p}\\
      & \sum_{k:(l,k) \in \mathcal{L}} q_{lk, \phi} + {q}^\text{inj}_{l,\phi}  = 0, \quad \forall l\in\mathcal{N}, \phi \in \{a,b,c\} \label{eq:power_balance_q}\\
   & {v}^{\text{min}} \leq v_{n,\phi} \leq {v}^{\text{max}}, \quad \forall n\in\mathcal{N}, \phi \in \{a,b,c\}. \label{eq:voltage_const}
\end{align}
\end{subequations}

These constraints enforce both the grid's operational requirements and phase consistency. Eq.~\eqref{eq:v_model} expresses the nodal voltage using the Lin3DistFlow model. Eqs.~\eqref{eq:p_inj_cons} and \eqref{eq:q_inj_cons} are the per-phase limits for the active ($\bar{p}_{n,\phi}$) and reactive ($\bar{q}_{n,\phi}$) power injections. The binary variable $\xi_{n,\phi}$ ensures that these constraints are only active when the phase is present at a bus. Eq.~\ref{eq:pf_constraint} ensures that the reactive power injection per phase is less than the active power injection. Eqs.~\eqref{eq:psum_inj_cons} and \eqref{eq:qsum_inj_cons} are the limits on the total injection per node, where $\hat{p}_{n,a}, \hat{p}_{n,b}, \hat{p}_{n,c}$ and $\hat{q}_{n,a}, \hat{q}_{n,b}, \hat{q}_{n,c}$ refer to the active and reactive power demands.
Eq.~\eqref{eq:phase_constraint} formalizes the phase constraint, ensuring that a single-phase line remains single-phase, as no new phases are added. Here, $|\mathcal{P}_n|$ denotes the number of phases at a given bus. This constraint is essential to prevent inadvertent phase expansion and to maintain phase consistency throughout the network. Eq.~\eqref{eq:phase_consistency} expresses the phase consistency constraint, i.e., downstream buses cannot have a new phase assigned, where $m_\text{child}$ is defined as in Section~\ref{sec:phase_consistency}. Eqs.~\eqref{eq:power_balance_p} and \eqref{eq:power_balance_q} refer to the power balance constraint for active and reactive powers. Eq.~\eqref{eq:voltage_const} ensures that voltage magnitudes are within operational bounds $[{v}^{\text{min}},~ {v}^{\text{max}}]$.

\section{Numerical Simulations}
\label{sec:numerical_sim}
We numerically validate the proposed algorithm on the IEEE-13, IEEE-37, and IEEE-123 benchmark systems. First, we present a detailed analysis using the IEEE-13 system, followed by results for the IEEE-37 and IEEE-123 networks. 
\begin{figure}[b]
    \centering
    \vspace{-2em}
    \subfloat[Original IEEE-13 network.]{\includegraphics[width=0.5\linewidth]{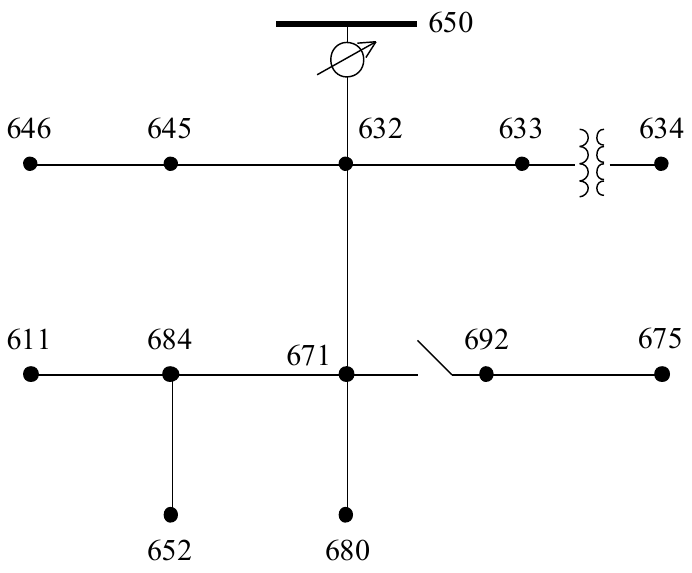}
    \label{fig:ieee13}}
    \subfloat[Equivalent phase representation.]{\includegraphics[width=0.55\linewidth]{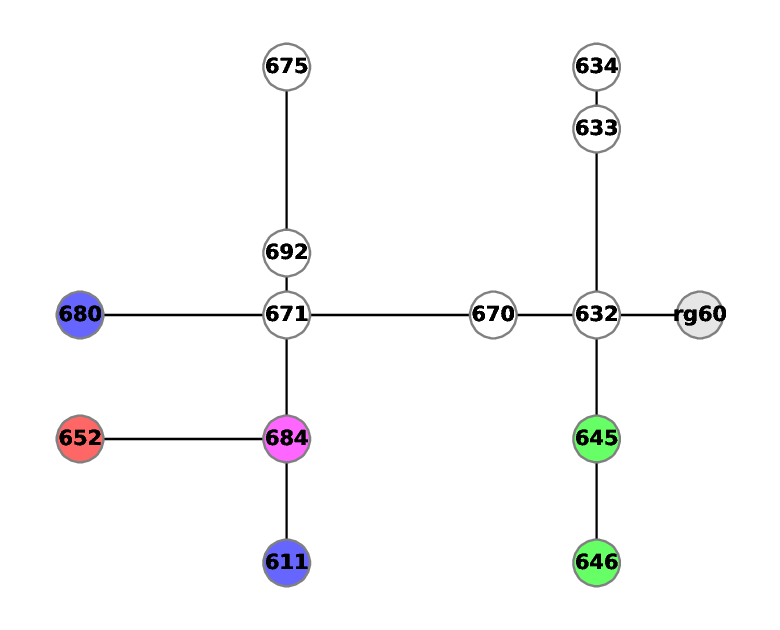}
    \label{fig:equivalent_case}}
    \caption{IEEE 13-bus benchmark network (base case). For the equivalent representation, the nodes are color-coded as follows: Red, Green, and Blue represent phases A, B, and C, respectively. Combinations of two or more colors indicate multiple phases; for example, Red + Blue = Purple means the presence of phases A and C, Blue + Green = Cyan means phases B and C, and Red + Green + Blue = White indicates the presence of all phases.}
    \label{fig:ieee13network}
\end{figure}
\begin{table}[b!]
    \caption{Spot load data.}
    \centering
    \begin{tabular}{|c||c|c|c|}
        \hline
         \bf Node & Phase -- A ($\hat{p}_{n,a}$) &  Phase -- B ($\hat{p}_{n,b}$) &  Phase -- C ($\hat{p}_{n,c}$) \\
                    & [kW / kVAr] & [kW / kVAr]  & [kW / kVAr] \\
         \hline
         670        & 17 / 10	    & 66 / 38	    & 117 / 68\\
         634        & 160 / 110	& 120 / 90	& 120 / 90  \\
         645	   & 0 / 0	    & 170 / 125	& 0 / 0\\
         646        & 0 / 0	    & 230 / 132	& 0 / 0 \\
         652        & 128 / 86	& 0 / 0	    & 0 / 0   \\
         671        & 385 / 220	& 385 / 220	& 385 / 220 \\
         675        & 485 / 190	& 68 / 60	    & 290 / 212 \\
         692        & 0 / 0	    & 0 / 0	    & 170 / 151 \\
         611        & 0 / 0	    & 0 / 0	    & 170 / 80 \\
         \hline \hline
         Total      & 1175 / 616	& 1039 / 665	& 1252 / 821\\
         \hline
    \end{tabular}
    \label{tab:ieee_load_data}
\end{table}

\begin{figure*}[!htbp]
    \vspace{-1em}
    \centering
    \subfloat[Case 1.]{\includegraphics[width=0.3\linewidth]{Figures/13Bus_basecase.eps}
    \label{fig:basecase}}
    \subfloat[Case 2.]{\includegraphics[width=0.3\linewidth]{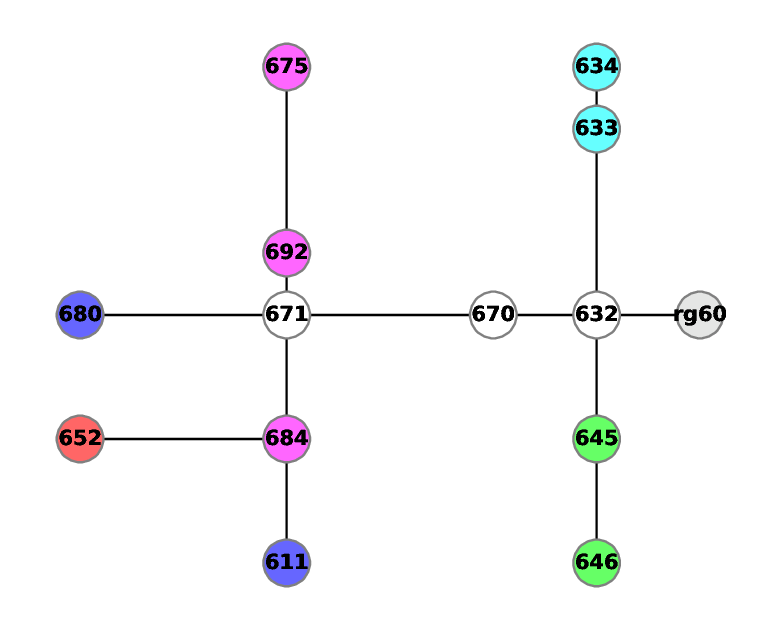}
    \label{fig:case1}}
    \subfloat[Case 3.]{\includegraphics[width=0.3\linewidth]{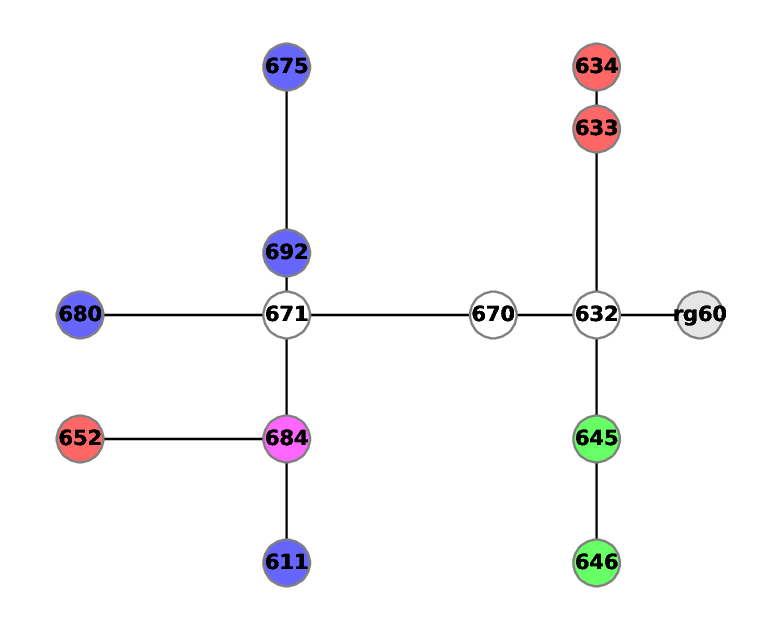}
    \label{fig:case2}}
    \caption{Optimized phase allocation is evaluated across three cases: Case 1, Case 2, and Case 3 correspond to the per-phase capacity to be equal to, double, and triple the per-phase spot load, respectively.}
    \label{fig:optimized_phase}
\end{figure*}
\begin{figure}[!htbp]
    \centering
    \subfloat[Case 1 (Per phase capacity equal to per phase spot load).]{\includegraphics[width=\linewidth]{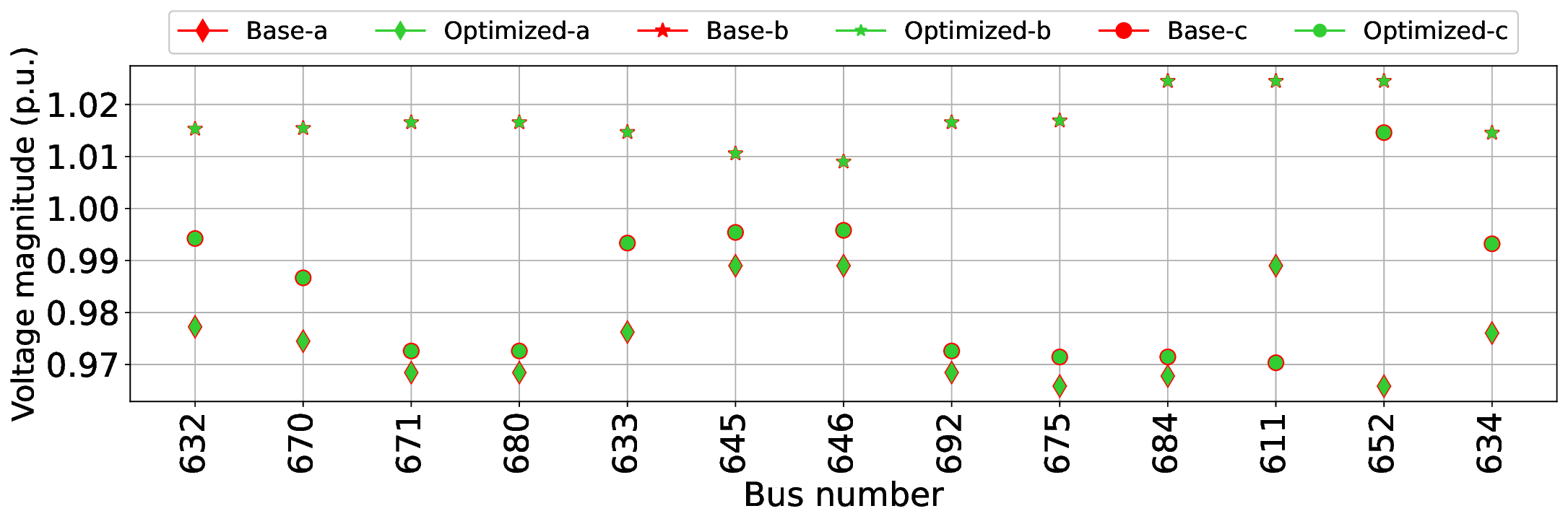}
    \label{fig:basecase_v}}\\
    \subfloat[Case 2 (Per phase capacity is double the per phase spot load).]{\includegraphics[width=\linewidth]{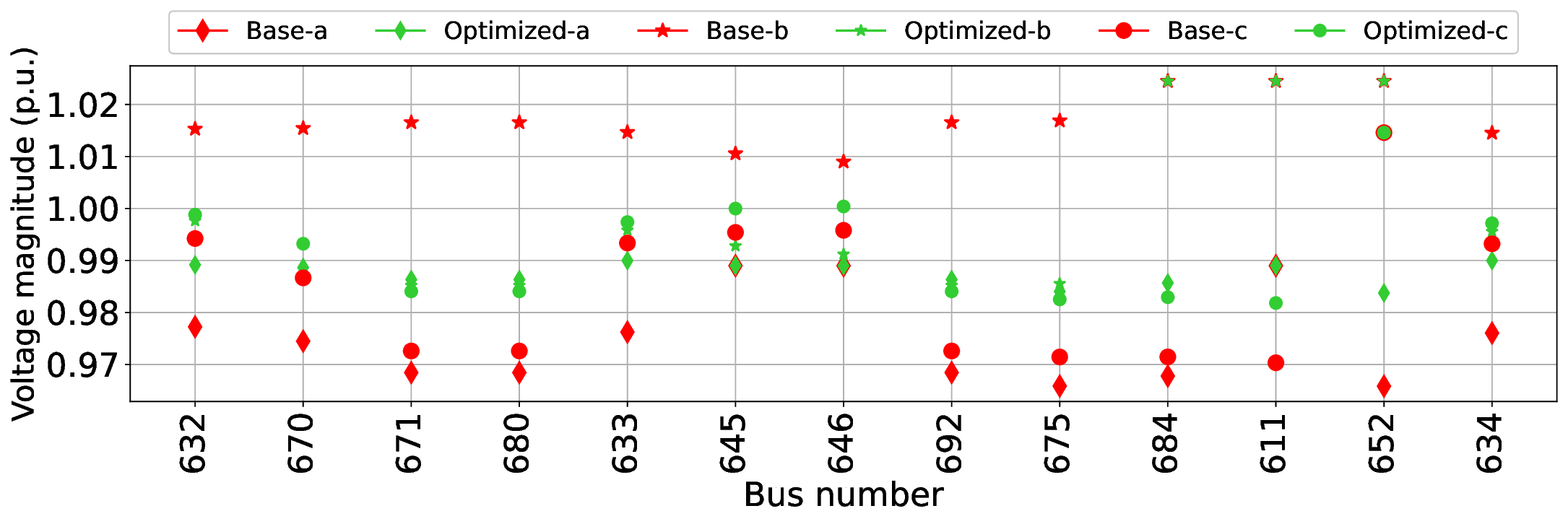}
    \label{fig:case1_v}}\\
    \subfloat[Case 3 (Per phase capacity is triple the per phase spot load).]{\includegraphics[width=\linewidth]{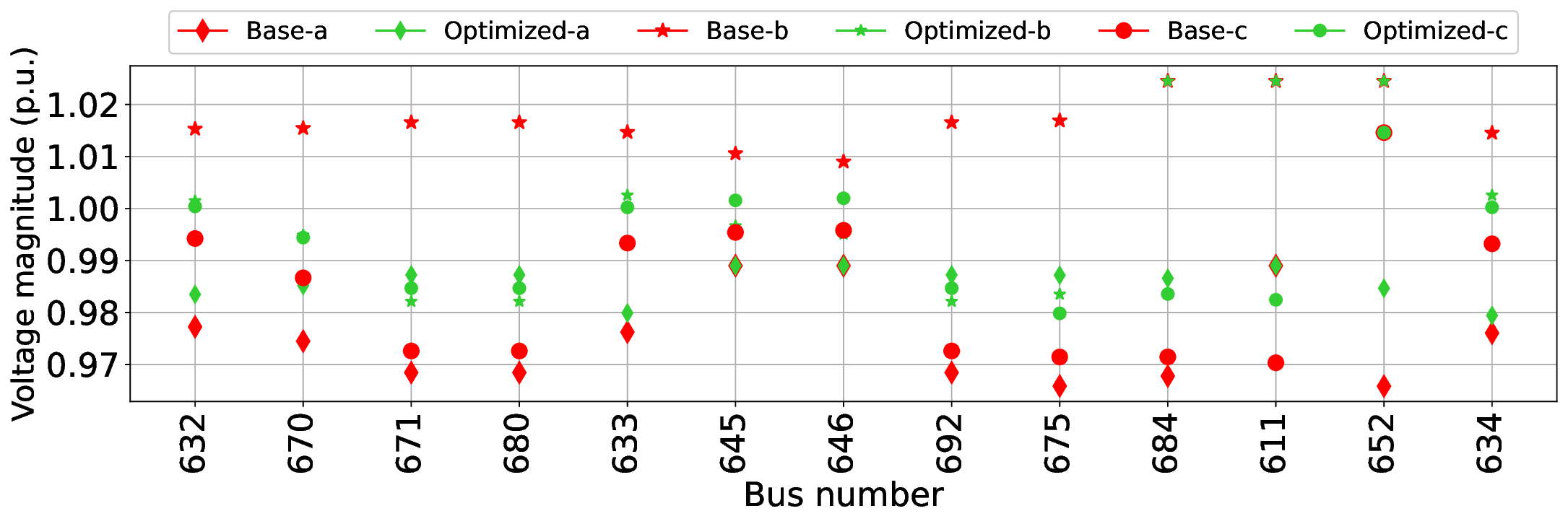}
    \label{fig:case2_v}}
    \caption{Voltage magnitudes for Cases 1, 2 and 3.}
    \label{fig:optimized_voltage}
\end{figure}
The IEEE-13 base network is shown in Fig.~\ref{fig:ieee13} and its equivalent phase type representation is shown in Fig.~\ref{fig:equivalent_case}, where the network parameters are obtained from \cite{IEEE_test}. The additional nodes '670' and 'rg60' in Fig.~\ref{fig:equivalent_case} are introduced as per the OpenDSS file. The active and reactive spot load per phase, denoted as $\hat{p}_{n,a}, \hat{p}_{n,b}, \hat{p}_{n,c}$ for a particular node $n$, is listed in Table~\ref{tab:ieee_load_data}.
For validation, we simulate the following three cases:
\begin{itemize}
    \item \textbf{Case 1:} We run the phase balancing algorithm with the nodal power injection limits in accordance with spot load data, i.e., $\bar{p}_{n,\phi} = \hat{p}_{n,\phi}$ and $\bar{q}_{n,\phi} = \hat{q}_{n,\phi}$.
    \item \textbf{Case 2:} In this case, we assume that nodes can host double the amount of spot load data, i.e., $\bar{p}_{n,\phi} = 2\times \hat{p}_{n,\phi}$ and $\bar{q}_{n,\phi} = 2\times \hat{q}_{n,\phi}$.
    \item \textbf{Case 3:} In this case, we assume that nodes can host three times the amount of spot load data, i.e., $\bar{p}_{n,\phi} = 3\times \hat{p}_{n,\phi}$ and $\bar{q}_{n,\phi} = 3\times \hat{q}_{n,\phi}$.
\end{itemize}
Cases 2 and 3 are relevant because networks are often over-sized during the design stage, allowing this extra capacity to help mitigate phase unbalance through phase swapping. The optimized phases for all three cases are shown in Fig.~\ref{fig:optimized_phase}.

The optimized phases for the three cases are shown in Fig.~\ref{fig:optimized_phase}, with nodes color-coded as described in the caption. Notably, there are significant differences in phase assignments between Cases 1, 2, and 3. The optimized phase of Case 1 is the same as in the base case (Fig.~\ref{fig:equivalent_case}).  Many three-phase nodes in case 1 were converted to two-phase nodes in Case 2 and to single-phase nodes in Case 3. This shift is due to an increase in node capacity by factors of 2 and 3, respectively.

The corresponding voltage magnitudes for each case are shown in Fig.~\ref{fig:optimized_voltage}. The base case maintains the same voltage profile as the original network, while Cases 2 and 3 exhibit improved voltage balance. This improvement is evident from the narrower spread of voltage magnitudes per node, which is smaller compared to the base case. These improvements in voltage unbalance are achieved through phase adjustments at certain nodes, as shown in Figs.~\ref{fig:case1} and~\ref{fig:case2}. Table~\ref{tab:load_optimized} lists the changes in demand per phase (shown in bold). The demand adjustments at specific nodes improve the overall network balance.

\begin{table*}[t]
    \renewcommand{\arraystretch}{0.8}
    \caption{Spot load reassignment under Cases 1, 2 and 3.}
    \centering
    \resizebox{\textwidth}{!}{%
    \begin{tabular}{|c||c|c|c||c|c|c||c|c|c|}
        \hline
        & \multicolumn{3}{c||}{\bf Case 1} & \multicolumn{3}{c||}{\bf Case 2} & \multicolumn{3}{c|}{\bf Case 3} \\
        \hline
         \bf Node & Phase -- A &  Phase -- B &  Phase -- C & Phase -- A &  Phase -- B &  Phase -- C & Phase -- A &  Phase -- B &  Phase -- C\\
                    & [kW / kVAr] & [kW/kVAr]  & [kW / kVAr] & [kW / kVAr] & [kW / kVAr]  & [kW / kVAr] & [kW / kVAr] & [kW / kVAr]  & [kW / kVAr]\\
         \hline
         670        & 17 / 10	    & 66 / 38	    & 117 / 68   &  \bf 0 / 0     & \bf 24.7 / 24.7    & \bf 175.3 / 91.3   & \bf 15.6 / 15.6     & \bf 184.4 / 100.4    & \bf 109.1 / 109.1  \\
         634        & 160 / 110	& 120 / 90	& 120 / 90  & \bf 0 / 0    & \bf 160 / 160   & \bf 240 / 130 & \bf 400 / 290    & \bf 0 / 0  & \bf 0 / 0\\
         645         & 0 / 0	    & 170 / 125	& 0 / 0  & 0 / 0	    & 170 / 125	& 0 / 0   & 0 / 0	    & 170 / 125	& 0 / 0 \\
         646        & 0 / 0	    & 230 / 132	& 0 / 0  & 0 / 0	    & 230 / 132	& 0 / 0  & 0 / 0	    & 230 / 132	& 0 / 0 \\
         652        & 128 / 86	& 0 / 0	    & 0 / 0    & 128 / 86	& 0 / 0	    & 0 / 0     & 128 / 86	& 0 / 0	    & 0 / 0   \\
         671        & 385 / 220	& 385 / 220	& 385 / 220 & \bf 0 / 0    & \bf 770 / 440   & \bf 385 / 220 & \bf 1045.9 / 550.9   & \bf 0 / 0   & \bf 0 / 0\\
         675        & 485 / 190	& 68 / 60	    & 290 / 212  & \bf 364.6 / 364.6    & \bf 0 / 0   & \bf 478.4 / 97.4 & \bf 0 / 0    & \bf 0 / 0   & \bf 843 / 462 \\
         692        & 0 / 0	    & 0 / 0	    & 170 / 151 & 0 / 0	    & 0 / 0	    & 170 / 151  & 0 / 0	    & 0 / 0	    & 170 / 151 \\
         611   & 0 / 0	    & 0 / 0	    & 170 / 80  & 0 / 0	    & 0 / 0	    & 170 / 80  & 0 / 0	    & 0 / 0	    & 170 / 80 \\
         \hline \hline
         Total     & 1175 / 616	& 1039 / 665	& 1252 / 821 & 492.6 / 450.6  & 1354.7 / 881.7 & 1618.7 / 769.7 & 1589.5 / 942.5  & 584.4 / 357.4 & 1292.1 / 802.1\\
         \hline
    \end{tabular}}
    \label{tab:load_optimized}
\end{table*}
\begin{table}[!htbp]
    \centering
    \caption{Comparison of Voltage Unbalances for IEEE benchmark networks.}
    \begin{tabular}{|c|c|c|c|c|}
        \hline
         \bf Network & \multicolumn{4}{c|}{\bf Unbalance Metric} \\
         \hline
                    & Base case & Case 1 & Case 2 & Case 3\\
         \hline 
         IEEE-13    & 0.66 & 0.66 & 0.22  & 0.29 \\
         \hline
         IEEE-37    & 1.63 & 1.63 & 1.55 & 1.48 \\
         \hline
         IEEE-123   & 2.92 & 2.92 & 1.93 & 1.81 \\
         \hline
    \end{tabular}
    \label{tab:other_testcases}
\end{table}

\begin{table}[!htbp]
    \centering
    \caption{Solving Time.}
    \begin{tabular}{|c|c|c|}
        \hline
        IEEE-13 & IEEE-37 & IEEE-123 \\
         \hline 
         0.1 sec. & 3.7 sec. & 5.0 sec. \\
         \hline
    \end{tabular}
    \label{tab:computation_time}
\end{table}

We also applied the proposed phase allocation problem on the IEEE-37 and IEEE-123 systems. As summarized in Table~\ref{tab:other_testcases}, the results are evaluated using a voltage unbalance metric that is defined as the sum of the difference between the magnitudes of the phase voltage and the mean voltage at each node, that is, $(\sum_{n\in\mathcal{N}}\sum_{\phi, \in \{a,b,c\}} |\underline{v}_{n} - v_{n,\phi}|)$. From the table, it is evident that the optimized cases (Cases 2 and 3) achieve a better voltage balance compared to the base case. Compared to Case 1, Case 2 improves the unbalance metric by 67\%, 5\% and 34\% for IEEE-13, IEEE-37 and IEEE-123, respectively, and Case 3 improves the unbalance metric by 56\%, 9\% and 38\%, respectively.
Case 3 for IEEE-13 performs worse than Case 2, highlighting that minimizing the number of phases is not always advantageous. The effectiveness of this minimization depends on the weight parameter $\alpha$ in \eqref{eq:obj}.

Furthermore, Table~\ref{tab:computation_time} compares the solving time for the phase balancing problem that is run on an Apple Macbook M2 Pro, 16 GB memory. As seen, the proposed scheme is quite fast and scaled very well with an increasing number of nodes.

\vspace{-1em}
\section{Conclusions}
\label{sec:conclusion}
This work proposes an approach for the phase allocation problem in power distribution networks. The algorithm is formulated as a mixed integer linear program (MILP), where phase decisions for each node are represented by binary variables. The formulation leverages the linearized DistFlow approximation to enable a linear representation. A phase consistency constraint is introduced to ensure that downstream phase configurations align with upstream configurations.

The proposed approach was validated on several IEEE benchmark networks. First, the approach was applied to the IEEE-13 base case, where the per-phase capacity at each node remained unchanged. In this case, the phase configuration remained the same as in the original test case. Additionally, two scenarios were considered where the per-phase capacity was doubled and tripled. In these cases, more balanced voltage profiles were observed in comparison to the base case. 

Finally, the algorithm was applied to the IEEE-37 and IEEE-123 networks, where similar behavior was observed: our approach achieved more balanced voltage magnitudes when the per-phase node capacities were increased.

Future work will extend this scheme to the AC power flow model, leading to a mixed-integer non-linear problem that requires more sophisticated solution approaches.
\bibliographystyle{IEEEtran}
\bibliography{bibliography.bib}
\end{document}